\documentclass[nature,letterpaper,twocolumn,superscriptaddress,amsmath,amsfonts,amssymb,floatfix,preprintnumbers,citeautoscript]
{revtex4}
\pdfoutput=1
\usepackage[T1]{fontenc}\usepackage[latin1]{inputenc}
\usepackage{dcolumn,graphicx,color,booktabs} \graphicspath{{Images/}}
\usepackage{amsmath,bm,hvfloat}
\usepackage[outercaption]{sidecap}
\renewcommand{\figurename}{Fig.}
\makeatletter\renewcommand{\fnum@figure}[1]{\figurename\thefigure.}\makeatother
\definecolor{DarkBlue}{rgb}{0,0,0.5}

\begin{document} \pagestyle{plain}

\newcommand{\hsp}{\hspace{0.5cm}}
\newcommand{\vsp}{\hspace{0.6cm}}
\newcommand{\tsz}{\small  \linespread{1} \selectfont}

\title{Strong pairing at iron $3d_{xz,yz}$ orbitals in hole-doped BaFe$_2$As$_2$}

\author{D.\,V.\,Evtushinsky}
\affiliation{Institute for Solid State Research, IFW Dresden, P.\,O.\,Box 270116, D-01171 Dresden, Germany}
\author{V.\,B.\,Zabolotnyy}
\affiliation{Institute for Solid State Research, IFW Dresden, P.\,O.\,Box 270116, D-01171 Dresden, Germany}
\author{T.\,K.\,Kim}
\affiliation{Institute for Solid State Research, IFW Dresden, P.\,O.\,Box 270116, D-01171 Dresden, Germany}
\affiliation{Diamond Light Source Ltd., Didcot, Oxfordshire, OX11 0DE, United Kingdom}
\author{A.\,A.\,Kordyuk}
\affiliation{Institute for Solid State Research, IFW Dresden, P.\,O.\,Box 270116, D-01171 Dresden, Germany}
\affiliation{Institute of Metal Physics of National Academy of Sciences of Ukraine, 03142 Kyiv, Ukraine}
\author{A.\,N.~Yaresko}
\affiliation{Max-Planck-Institute for Solid State Research, Heisenbergstrasse 1, D-70569 Stuttgart, Germany}
\author{J.\,Maletz}\author{S.\,Aswartham}\author{S.\,Wurmehl}
\affiliation{Institute for Solid State Research, IFW Dresden, P.\,O.\,Box 270116, D-01171 Dresden, Germany}
\author{A.\,V.~Boris}
\affiliation{Max-Planck-Institute for Solid State Research, Heisenbergstrasse 1, D-70569 Stuttgart, Germany}
%\affiliation{Department of Physics, Loughborough University, Loughborough, LE11 3TU, United Kingdom}
\author{D.\,L.\,Sun}\author{C.\,T.~Lin}
\affiliation{Max-Planck-Institute for Solid State Research, Heisenbergstrasse 1, D-70569 Stuttgart, Germany}
\author{B.\,Shen}
\affiliation{Institute of Physics, Chinese Academy of Sciences, Beijing 100190, China}
\author{H.\,H.\,Wen}
\affiliation{National Laboratory of Solid State Microstructures and Department of Physics, Nanjing University, Nanjing 210093, China}
\author{A.\,Varykhalov}\author{R.\,Follath}
\affiliation{BESSY GmbH, Albert-Einstein-Strasse 15, 12489 Berlin, Germany}
\author{B.\,B\"{u}chner}
\affiliation{Institute for Solid State Research, IFW Dresden, P.\,O.\,Box 270116, D-01171 Dresden, Germany}
\author{S.\,V.\,Borisenko}
\affiliation{Institute for Solid State Research, IFW Dresden, P.\,O.\,Box 270116, D-01171 Dresden, Germany}

\begin{abstract}
\noindent
\end{abstract}

% Experimental input is immense owing to large diversity of available compounds.

%\pacs{74.25.Fy, 74.25.Jb, 79.60.-i, 71.20.-b}

\maketitle

\begin{figure*}[]
\flushright
\begin{minipage}[l]{0.68\linewidth}
\includegraphics[width=\textwidth]{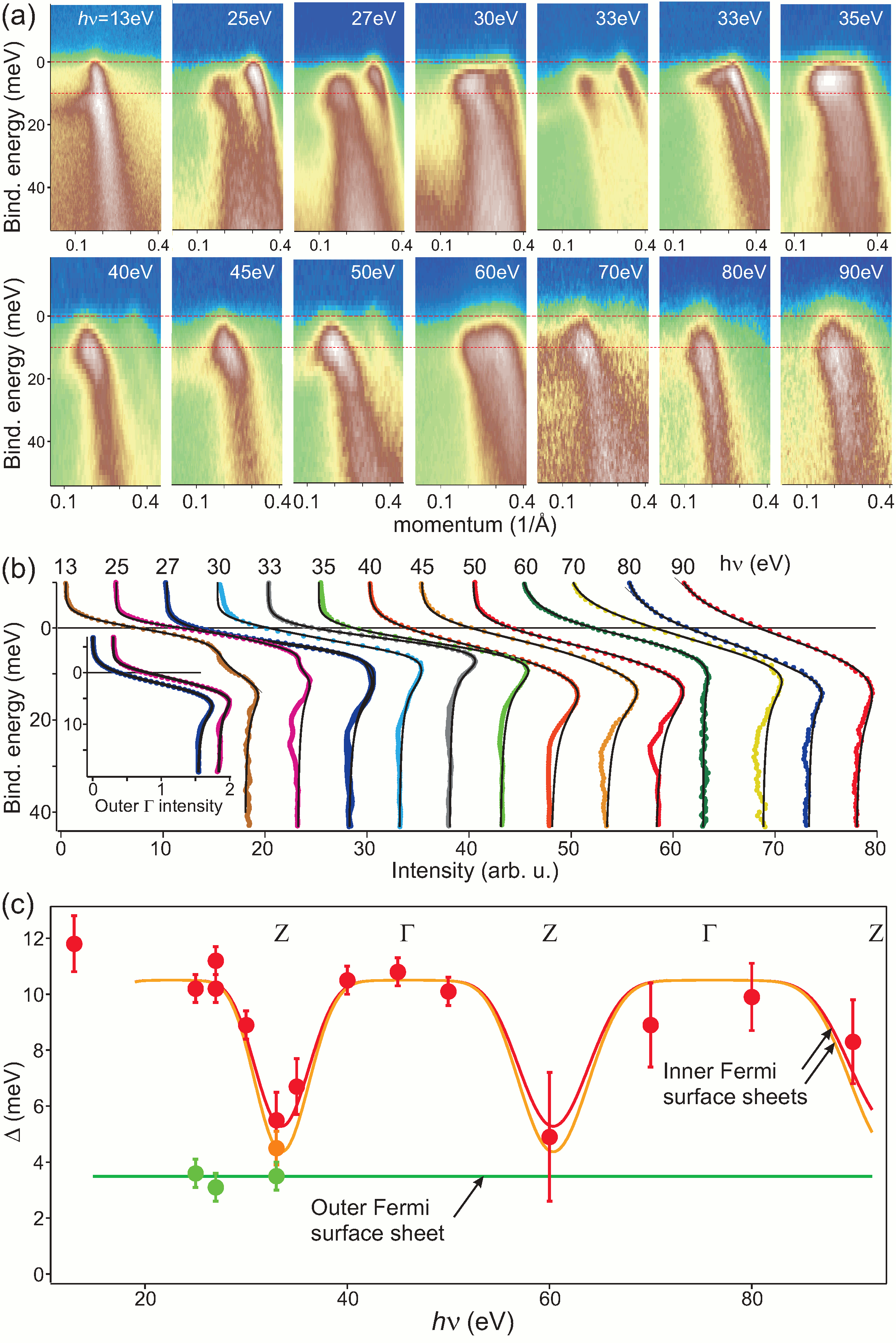}
%\caption{default}
%\label{fig:figure1}
\end{minipage}
\hspace{0.015\linewidth}
%\hspace{-0.2\linewidth}
\begin{minipage}[r]{0.28\linewidth}
\caption{$k_z$-dependent superconducting gap at hole-like Fermi surface sheets centered at the ($k_x=0$, $k_y=0$) point in optimally doped Ba$_{1-x}$K$_x$Fe$_2$As$_2$. (a) Energy-momentum cut, passing through the ($k_x=0$, $k_y=0$)-point and capturing spectrum around Fermi crossings of $\Gamma$ FS sheets, recorded at $h\nu$'s ranging from 13 to 93\,eV. As discussed in the main text, the two nearly degenerate inner $\Gamma$-barrels are often not resolved. All three $\Gamma$-barrels are clearly resolved at excitation energies around the value of $h\nu=33.5$\,eV. Red dashed lines are guides to the eye, located at 0 and 10\,meV binding energy. (b) Integrated energy distribution curves (IEDC) fitted to the Dynes function. (c) Values of the superconducting gaps at $\Gamma$-barrels as a function of $h\nu$. Underlying fitting curves are approximated by simple formula derived for a free-electron-like final state and inner potential of 7.6\,eV. Note the large flat regions in $k_z$-dependence of $\Delta_{\rm large}$ in between of cusps at $h\nu=33.5$, 60 and 93\,eV, corresponding to the variation of intensity in Fig.~\ref{kz}~(b). Taking into account that $k_z$-dependence of the gap is observed only for the inner $\Gamma$-barrels and there it possesses large flat regions, the two-gap model still can be sufficient for satisfactory interpretation of many experimental results.} \vspace{4cm}
\label{gap_kz}
\end{minipage}

\end{figure*}

\textbf{Among numerous hypotheses, recently proposed to explain superconductivity in iron-based superconductors \cite{Kivelson, Kuroki, Chubukov, Johnston, Chubukov2, Kotliar, Imai, Hardy, Ding_nesting}, many consider Fermi surface (FS) nesting \cite{Johnston, Kuroki, Ding_nesting, Castellan} and dimensionality \cite{Johnston, Hardy} as important contributors. Precise determination of the electronic spectrum and its modification by superconductivity, crucial for further theoretical advance, were hindered by a rich structure of the FS \cite{VolodyaNature, YareskoPRB, FengPRL, DingNat, BorisenkoLiFeAs, Ding_FeTeSe, Feng_AFeSe}. Here, using the angle-resolved photoemission spectroscopy (ARPES) with resolution of all three components of electron momentum and electronic states symmetry, we disentangle the electronic structure of hole-doped BaFe$_2$As$_2$, and show that nesting and dimensionality of FS sheets have no immediate relation to the superconducting pairing. Alternatively a clear correlation between the orbital character of the electronic states and their propensity to superconductivity is observed: the magnitude of the superconducting gap maximizes at 10.5\,meV exclusively for iron $\mathbf{3d_{xz,yz}}$ orbitals, while for others drops to 3.5\,meV. Presented results reveal similarities of electronic response to superconducting and magneto-structural transitions \cite{Shen_SDW, Braden}, implying that relation between these two phases is more intimate than just competition for FS, and demonstrate importance of orbital physics in iron superconductors.}

%At the stage when pure theory is not ready to provide clues for understanding the nature of the high temperature superconductivity in iron-based high-$T_{\rm c}$ materials, it seems perfectly sensible to search for phenomenological tendencies, to push ourselves closer to the solution of the problem.

There are several experimentally established tendencies, which are followed by many representatives of iron-based family with highest $T_{\rm \text{c}}$: presence of the electronic states with large difference in superconducting gap magnitude, often a two-gap behavior \cite{EvtushinskyNJP, Hardy, BorisPRL, Pramanik}, gap-to-$T_{\rm \text{c}}$ ratios much higher that the universal BCS value \cite{Hardy, EvtushinskyNJP, Bogoliubons}, correlation of $T_{\rm \text{c}}$ with anion height \cite{Okabe}. The phenomenology requires a large amount of input data and therefore is usually applied to a set of materials. Large diversity of electronic states at the Fermi level, found in Ba$_{1-x}$K$_x$Fe$_2$As$_2$ (BKFA), originally was an obstacle on the way to complete understanding of the underlying electronic structure \cite{VolodyaNature, DingEPL}. At closer look such variety of electronic states turned out to be a blessing, allowing for detailed studies of the response of different states to the superconducting transition within the same material.

The superconducting gap in BKFA was studied by means of various experimental techniques \cite{BorisPRL, EvtushinskyNJP}, and vast majority of the results can be interpreted in terms of presence of comparable amount of electronic states gapped with a large gap ($\Delta_{\rm large}=$10--11\,meV) and with a small gap ($\Delta_{\rm small}<4$\,meV). The in-plane momentum dependence of the superconducting gap, determined in early ARPES studies, is the following: the large gap is located on all parts of the FS except for the outer hole-like FS sheet around $\Gamma$-point \cite{DingEPL, EvtushinskyPRB, GBarrel}. Though the crystal structure of iron-based superconductors of interest is layered, and electronic states at the Fermi level are formed mainly by the atomic orbitals of the iron planes, studying the dependence of the electronic spectrum on the out-of-plane momentum leads to quite interesting observations.

The resolution of the electronic spectra along the out-of-plane momentum is achieved in photoemission experiments via the variation of the energy of the incoming photons. Fig.~\ref{gap_kz}(a) shows an energy-momentum cut capturing the Fermi crossings of the inner and outer $\Gamma$ FS sheets, imaged with excitation energies in the range from 13 to 90\,eV. The values of the superconducting gap were determined from the fit of the integrated energy distribution curves (IEDC) to the Dynes function \cite{Dynes, EvtushinskyPRB}. IEDCs and fitting curves are shown in the panel (b). Upon analysis of ARPES spectra of the BKFA one unavoidably encounters a problem of additional, presumably surface-related, contributions to the signal. This issue is discussed in the Supplementary Materials, here we only note that fitting data allowed us to single out the relevant values for the gap. The derived $h\nu$ dependence is shown in Fig.~~\ref{gap_kz}~(c). The gap on the inner $\Gamma$-barrel varies from maximal value of $10.5\pm0.5$\,meV, observable for $h\nu$=13, 25, 45\,eV, to the minimal value of 4.5\,meV at $h\nu$=33.5\,eV. At photon energies close to 33\,eV it is possible to clearly resolve two components of the inner $\Gamma$-barrel as separate features in the spectra [see Fig.~\ref{gap_kz}~(a) and Fig.~S\ref{sigma_gap}]; at 33\,eV the gaps for the $\Gamma$-barrels are: 3.5\,meV for the outer one, 4.5\,meV for the second and 5.5\,meV for the innermost.

\begin{figure*}[]
\flushright
\begin{minipage}[l]{0.74\linewidth}
\includegraphics[width=\textwidth]{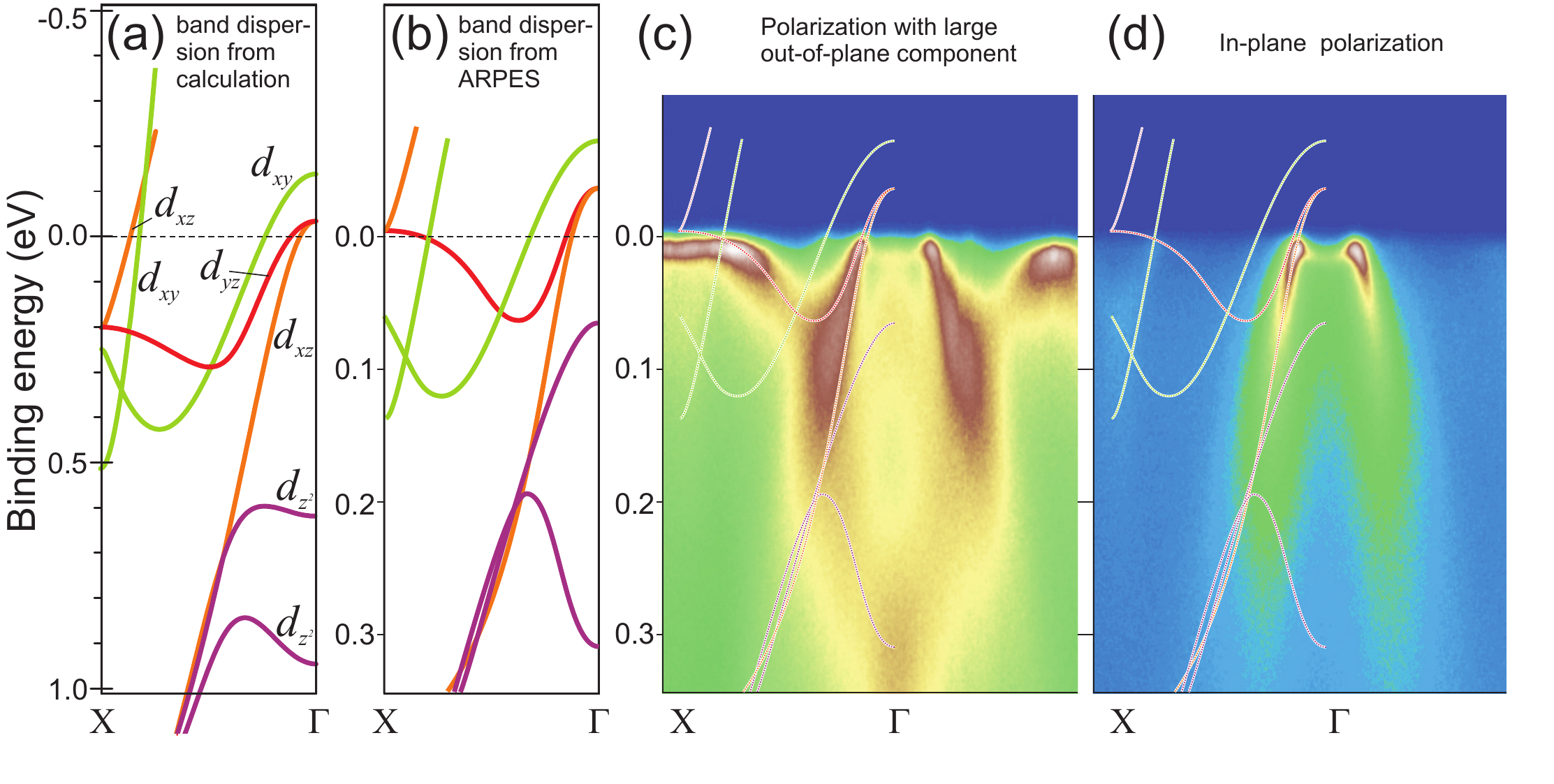}
%\caption{default}
%\label{fig:figure1}
\end{minipage}
\hspace{0.015\linewidth}
%\hspace{-0.2\linewidth}
\begin{minipage}[r]{0.22\linewidth}
\vspace{0.6cm}
\caption{Comparison of the calculated band dispersion, to the band dispersion, extracted from ARPES data. (a) The calculated band dispersion. (b) The band dispersion, extracted from ARPES data, presented in panels (c,d). (c) Energy-momentum cut, passing through the ($k_x=0$, $k_y=0$) point, recorded at 70\,eV using horizontal light polarization. (d) Same cut, recorded with vertical polarization.}
\label{ARPES_calc}
\end{minipage}

\end{figure*}

$\Delta_{\rm large}(h\nu)$ is quasiperiodical, shows large flat regions with magnitude close to maximal, and rather steep decreases to the minimum and steep increases back to flats, which is particularly well illustrated by the cusp around 33.5\,eV. The intensity distribution in the FS map (Fig.~S\ref{kz}), also shows rather fast variations at 33.5, 60 and 93\,eV, and regions with rather smooth variation of the signal in between. Interestingly, there is a rather transparent connection of such $h\nu$-dependence of the photoemission signal to the $k_z$-dependence of the electronic states at the Fermi level in the band structure calculation: as seen in Fig.~S\ref{Yaresko}, there is a band at ($k_x=0$, $k_y=0$), which is far above the Fermi level at $k_z=0$ and comes down and starts to interact with two $d_{xz,yz}$-derived hole-like bands in a rather narrow region around $k_z=\pi$. The gap on the outer $\Gamma$-barrel, $\Delta_{\rm small}$, remains in the range of $3.5\pm0.5$\,meV for the whole range of different $k_z$ values [Fig.~~\ref{gap_kz}~(c)], i.e. is essentially $k_z$-independent. This behavior is inline with absence of $k_z$ dependence of the electronic states, forming the outer $\Gamma$-barrel, in the band structure calculations (see Fig.~S\ref{Yaresko}).

Next we identify the orbital composition of the electronic states at the Fermi level by matching the bands, observed in the ARPES spectra of BKFA, to the bands, obtained in the band structure calculations. The origin of the hole-like bands in the Brillouin zone (BZ) center is clear\,---\,the outermost barrel corresponds to the band formed prevalently of iron $3d_{xy}$ orbitals, while two inner barrels consist of combinations of $3d_{xz}$ and $3d_{yz}$ orbitals \cite{DingEPL, EvtushinskyPRB, FengPRL, Bogoliubons, DingNat}.The situation with the propeller-shaped bands, observed by ARPES at the BZ corner, seems to be more tricky\,---\,original FS, derived in band structure calculations, does not contain this feature \cite{VolodyaNature}. On the other hand, in the calculations there is a band, situated below the Fermi level, with dispersion rather similar to the one observed experimentally for the band supporting propeller FS sheets (see Fig.\,\ref{ARPES_calc} and Fig.~S\ref{Yaresko}). Therefore we employ an additional tool for identification of the electronic states\,--\,analysis of spectra, recorded with differently polarized incoming light. The grounds for such analysis stem from symmetry considerations\,---\,basing on the symmetry of the electronic wave function in the crystal and polarization direction of the incoming light, it is possible to show that the matrix element of photoemission is zero for certain experimental conditions.

Let us now consider the electronic states along $\Gamma$X line. Fig.\,\ref{ARPES_calc}\,(a) shows the results of calculations for this direction. We will now concentrate on the band, plotted in red; this band is composed of $d_{yz}$ orbitals \cite{Feng_BFAP}. Figs.\,\ref{ARPES_calc}\,(c,d) show ARPES spectra, recorded in the same $\Gamma$X direction with light polarizations perpendicular (c) and parallel (d) to the $\Gamma$X. One of the most prominent changes, observed upon switching polarization is complete disappearance of the band, which was the brightest in the panel (c) and possesses dispersion very similar to the just discussed red band from the calculations. Remarkably, the $d_{yz}$ orbital is odd with respect to the reflection in the plane containing $\Gamma$X and the direction to the detector, and therefore (due to mentioned symmetry reasons) should yield zero intensity when excited by the light polarized along $\Gamma$X. This implies that not only the $d_{yz}$ band in the calculation possesses the dispersion similar to the dispersion of the propeller bands, derived from ARPES, but also the strong polarization dependence, observed in ARPES, is fully consistent with the one, expected in theory. Panel (b) shows the calculated band dispersions, which are shifted in order to fit the ARPES spectra.

The last tile in the complete definition of the tree-dimensional electronic structure at the Fermi level is the gap at the propeller-like bands at the BZ corner. The variations of the gap magnitude on the propellers were not observed when changing $h\nu$, implying that the gap magnitude on the propeller remains $\sim10$\,meV for all values of $k_z$, in accord with previous reports \cite{FengPRL, DingNat} and with little $k_z$-dependence of the corresponding bands in the band structure calculations.

\begin{figure*}[]
\includegraphics[width=0.85\textwidth ]{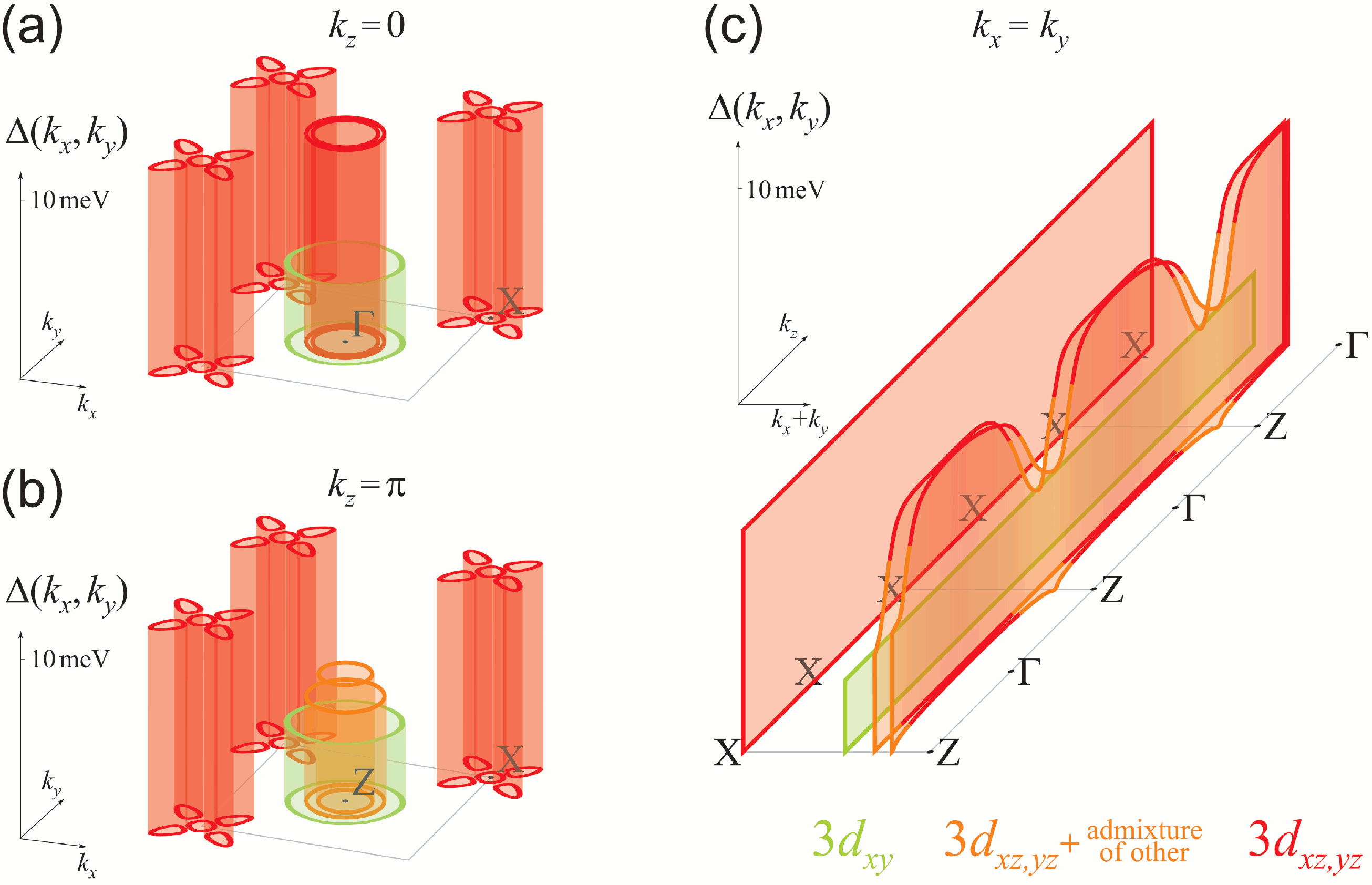}
\caption{Three-dimensional distribution of the superconducting gap and orbital composition of the electronic states at the Fermi level. (a) Distribution of the superconducting gap (plotted as height) and distribution of the orbital composition for the states at the Fermi level (shown in color: $d_{xz,yz}$\,---\,red, $d_{xy}$\,---\,green, $d_{xz,yz}$ with admixture of other orbitals\,---\,orange) as function of $k_x$ and $k_y$ at constant $k_z=0$; (b) the same, only for $k_z=\pi$; (c) same distributions as function of in-plane momentum, directed along BZ diagonal, and $k_z$. Note unambiguous correlation between the color and height, i.e. there is strong correlation between the orbital composition and superconducting gap magnitude.}
 \label{3D_gap}
\end{figure*}

Now, with exhaustive information on the distribution of the superconducting gap and orbital composition of the electronic states at the Fermi level at hand (Fig.\,\ref{3D_gap}), we see that there is a strong correlation of these two parameters. Namely, the gap is large for those and only for those states which originate purely from $d_{xz,yz}$ atomic orbitals. Indeed: (i) the gap is 3.5\,meV for the outer $\Gamma$-barrel, composed of $d_{xy}$ orbitals; (ii) the gap is 10\,meV for the propellers, composed of $d_{xz,yz}$ orbitals; (iii) for the inner $\Gamma$-barrels the gap reaches 10.5\,meV around $k_z=0$, where the electronic states originate from $d_{xz,yz}$, and drops to 5\,meV at $k_z=\pi$, where an admixture of $d_{3z^2-1}$ appears; (iv) additionally, in Supplementary Materials an evidence is presented for the 3\,meV gap at extra iron $3d_{3z^2-1}$ band, barely reaching the Fermi level.

The relation between the magnitude of the superconducting gap and orbital
composition of the electronic state, established here on the example of
Ba$_{1-x}$K$_{x}$Fe$_2$As$_2$, holds, at least partially, also for other
iron-based superconductors, BaFe$_2$As$_{2-2x}$P$_{2x}$, LiFeAs,
FeTe$_{1-x}$Se$_x$ \cite{Sergey_LiFeAs1, Ding_FeTeSe, Feng_BFAP_node, BFAP_node}, though a thorough analysis of this issue has not been performed. An indirect conformation for orbital dependence of the superconducting gap comes from ubiquitous observation of multigap behavior in iron-based superconductors \cite{EvtushinskyNJP, BorisPRL, Pramanik, Hardy} and a superconductivity-induced suppression of an absorption band \cite{BorisNatCom}.

The special role of iron $3d_{xz,yz}$ orbitals was also noticed in ARPES studies of the magneto-structural transition in the undoped and underdoped Ba-122 iron arsenides \cite{Shen_SDW}, implying that electronic states, which are affected most strongly by the magnetic ordering, appear to bear the largest gap in the superconducting state. This means that relation between superconductivity and magnetism in iron-based superconductors is more intimate than between phases just competing for the Fermi surface.

%Hyoung Joon Choi, David Roundy, Hong Sun, Marvin L. Cohen and Steven G. Louie, ``The origin of the anomalous superconducting properties of MgB2'', Nature \textbf{418}, 758 (2002).

In summary, the momentum distribution of the superconducting gap is substantially three-dimensional and rather nontrivial. Such momentum dependence of the superconducting gap is not predicted in simple models, where the pairing strength is determined by conventional Fermi surface nesting or dimensionality of Fermi surface sheets. Alternatively the correlation of the gap magnitude with the orbital composition of the electronic states takes place; in particular, the largest gap values were observed for iron $3d_{xz,yz}$ states.

%Though there still is a room for effects of magnetic ordering.

%\section{Summary}
%We have investigated the three-dimensional band structure and momentum dependence of the superconducting gap in Ba$_{1-x}$K$_{x}$Fe$_2$As$_2$ ($T_{\rm c}=38$\,K) by ARPES with variable incoming light. Observation of the quasi-periodic variation of photoemission signal with $h\nu$, which fits the model with free-electron final state, confirms possibility to probe electronic states at different $k_z$ separately in ARPES experiments on Ba$_{1-x}$K$_{x}$Fe$_2$As$_2$. Weak $k_z$-dispersion was observed for the inner $\Gamma$ FS sheets, while for the outer $\Gamma$-barrel and propeller bands $k_z$-dispersion is even weaker, below the accuracy of the present study. Noticeable variation of the superconducting gap with $k_z$ was found only for the inner $\Gamma$ barrels.

%All of five principal FS sheets\,---\,three hole-like $\Gamma$-barrels propeller's blades and shaft\,---\,are quasi two-dimensional.
%Excitation-energy-dependence of the photoemission signal, recorded in the wide range of $h\nu=13..90$\,eV, exhibits quasiperiodic character and can be fitted very well by a model, assuming free-electron-like final state, which implies that one indeed can probe electronic states at different $k_z$ values by changing $h\nu$.

\section*{Methods}

Measurements were carried out at the $1^3$-ARPES end station at BESSY synchrotron in Berlin (Helmholtz-Zentrum f\"{u}r Materialien und Energie) on the single crystals of optimally doped Ba$_{1-x}$K$_{x}$Fe$_2$As$_2$ with $T_{\rm c}=38$\,K \cite{Sun, BorisPRL, Luo}. All presented data are taken from optimally doped Ba$_{1-x}$K$_{x}$Fe$_2$As$_2$ except for the data in Fig.\,\ref{ARPES_calc}\,(c,d), which were recorded from the optimally doped Ba$_{1-x}$Na$_{x}$Fe$_2$As$_2$ with $T_{\rm c}=34$\,K \cite{Pramanik, 122KNa}.

%\section*{References}

\section*{Acknowledgements}
We thank A.\,V.\,Chubukov for helpful discussions, to R.\,H\"{u}bel and R.\,Sch\"{o}nfelder for technical support.

\clearpage

\setcounter{figure}{0}

\begin{center} \LARGE
\textbf{Supplementary materials}
\end{center}

\renewcommand{\figurename}{Fig. S}

\subsection{$k_z$-resolution in ARPES on $\text{Ba}_{1-x}$K$_x\text{Fe}_2\text{As}_2$}

In ARPES experiments one can naturally resolve the electronic spectrum along momentum components parallel to the sample surface, as $k_x$ and $k_y$ are conserved upon photoexcitation of electron from a crystal to vacuum. Variation of the energy of incoming photons, $h\nu$, may add the resolution of the third momentum component, $k_z$, if the nature of the final state is known. Within a simple model, assuming a free-electron-like final state, the dependence of $k_z$ on $h\nu$ is rather simple \cite{Dima, Huefner},
\begin{equation}
k_z\sim\sqrt{h\nu + {\rm const}}.
\end{equation}
%There are objections on utilization free electron approximation for the final state
As was found in previous studies, the FS of BKFA consists of a propeller-like structure near the Brillouin zone corner and roundish hole-like sheets at the $\Gamma$-point \cite{VolodyaNature, VolodyaPhysC, EvtushinskyPRB}. A more precise inspection of the central region of the Brillouin zone have shown that the inner $\Gamma$-barrel is double-walled and there is another hole-like band \cite{FengPRL, Bogoliubons, DingNat}. Fig.~S1(a) shows the distribution of the photoemission intensity at the Fermi level, superimposed by the FS contours.  Fig.~S1(b) shows excitation energy dependence of the intensity distribution along the line, indicated by red dotted line on the map in panel (a). Clear quasiperiodicity of the photoemission signal is observed with increasing distance between equivalent points, inline with formula (1), which provides evidence for probing different $k_z$ at different $h\nu$s. Additionally the magnitude of the superconducting gap, derived from experimental data changes with same quasiperiodicity in $h\nu$ (Fig.~\ref{gap_kz}).

\begin{figure}[]
\includegraphics[width=\columnwidth ]{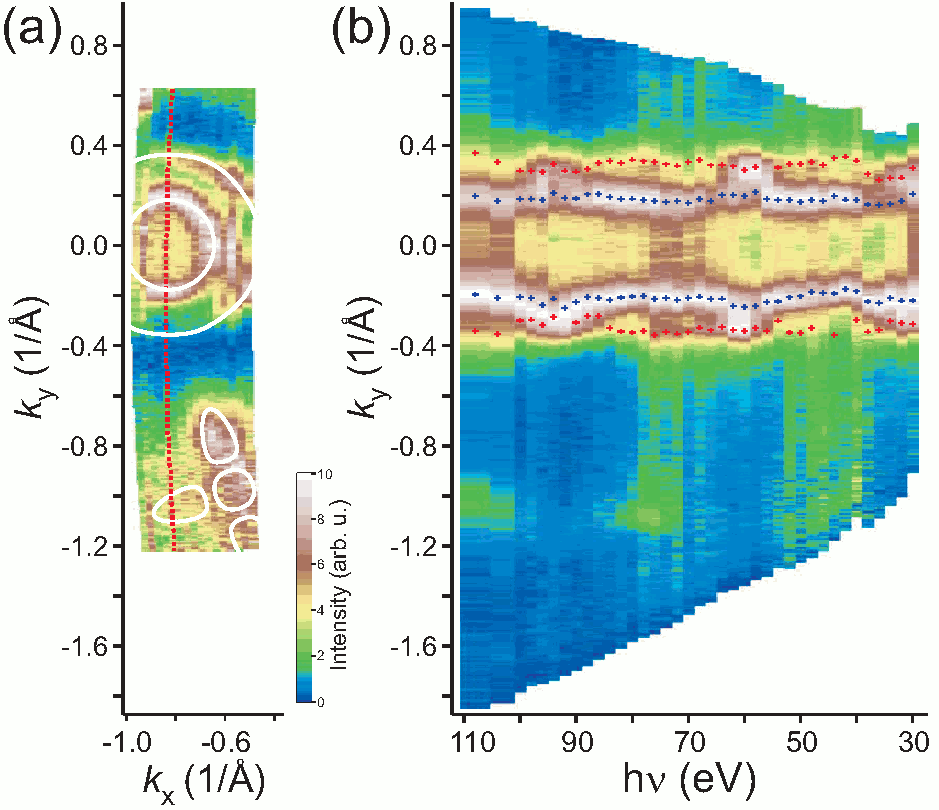}
\caption{ARPES with variable photon energies. (a) Fermi surface map of optimally doped Ba$_{1-x}$K$_{x}$Fe$_2$As$_2$ measured at $h\nu=50$\,eV. (b) Excitation energy map: intensity distribution along cut, indicated in panel (a) at the Fermi level (FL) was recorded using $h\nu$ in the range of 30--112\,eV.}
 \label{kz}
\end{figure}

\begin{figure*}[]
\includegraphics[width=\textwidth ]{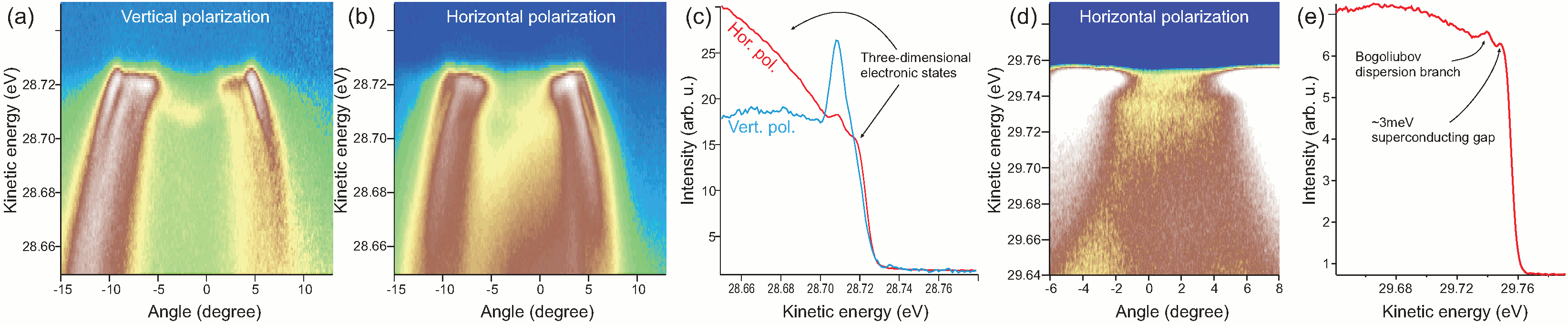}
\caption{Observation of the additional spectral weight from iron $3d_{3z^2-1}$ orbitals at the Fermi level and superconducting gap on it for optimally doped BKFA. (a) Energy-momentum cut, passing through the $(k_x=0,k_y=0)$-point, recorded with $h\nu=33$\,meV and vertical polarization of the incoming light. (b) The same cut, recorded with horizontal polarization. (c) EDCs directly from $(k_x=0,k_y=0)$-point recorded with horizontal and vertical polarizations. Comparison of EDCs recorded at different polarization suggests that spectral weight originating from three-dimensional electronic states, seen in spectrum as rather smeared intensity [see panel (b)] reaches the Fermi level. (d) Energy-momentum cut, passing through the $\Gamma$-point. (e) $\Gamma$-EDC reveals two peaks near the Fermi level: one, at binding energy about 17\,meV corresponds to the observed in this compound ``fusion of bogoliubons'' \cite{Bogoliubons}, while the one closer to the Fermi level corresponds to the superconducting gap with magnitude about 3\,meV.}
 \label{sigma_gap}
\end{figure*}

%The momentum distribution of the leading edge midpoint position, which is a simple spectroscopic representative of the gap \cite{kord_LEG, EvtushinskyPRB}, is shown in the Fig.~2(c), confirming the results, obtained from fit: the leading edge for the inner $\Gamma$-barrels is located at higher binding energies than the leading edge of the outer $\Gamma$-barrel.

\subsection{Spectral weight, related to $3d_{3z^2-1}$ electronic states at ($k_x=0,k_y=0$)}

Apart from the mentioned three bands, forming quasi two-dimensional hole-like FS sheets in the center of BZ, there is also a rather diffuse intensity below the Fermi level, centered at about 100\,meV binding energy [see e.g. Fig.~S\ref{sigma_gap}~(b,d), Fig.~\ref{ARPES_calc}~(c)]. In the band structure calculations there is a good candidate to account for this intensity\,---\,a band, formed by iron $3d_{3z^2-1}$ orbitals strongly hybridized with As $p_z$ states, see Fig.~S\ref{Yaresko}~(c), Fig.~\ref{ARPES_calc}. This band has been noticed in previous ARPES experiments on BKFA \cite{FengPRL} and rather extensively mapped for the case of KFe$_2$As$_2$ \cite{Yoshida_KFA}. Panels (a) and (b) of Fig.~S\ref{sigma_gap} show the same energy-momentum cut, passing through the $(k_x=0,k_y=0)$ point, recorded with $h\nu=33$\,eV and different polarizations of the incoming light. The electronic states of interest produce much photoemission intensity for the case of horizontal polarization [see panel (b)] and are almost invisible in the vertical polarization [panel (a)]. Panel (c) shows the EDCs for $(k_x=0,k_y=0)$ from panels (a) and (b). Not only the electronic states at higher binding energies yield more intensity when switching polarization from vertical to horizontal, but also some additional intensity appears at the Fermi level, as shown by arrows in panel (c), implying that at least some spectral weight, related to this $3d_{3z^2-1}$ band, reaches Fermi level due to self-energy broadening or/and crossing the Fermi level at some $k_z$ values. In panel (d) we show an energy-momentum cut, recorded at 1\,K with $h\nu=34$\,eV. One can notice two stripes close to the Fermi level at $(k_x=0,k_y=0)$: the one at higher binding energy is related to the fusion of Bogoliubov dispersion branches, observed before in BKFA \cite{Bogoliubons}, while the second stripe in closer vicinity to the Fermi level remained unnoticed before. In order to better visualize this spectral feature we present EDC from $(k_x=0,k_y=0)$ point in the panel (e). The arrows indicate the peaks\,---\,coming from the fusion of bogoliubons and the second newly detected peak. We propose that the second peak stems from a small superconducting gap opened at the discussed above electronic states, originating from a lower-lying band with $3d_{3z^2-1}$ character. From a fit to Dynes function we have determined that the magnitude of this superconducting gap is about 3\,meV.
Alternatively this new feature may originate from the spectral weight, related to the upper Bogoliubov dispersion branch. Though such interpretation does not explain the observed polarization dependence [Fig.~S\ref{sigma_gap}~(a,b)] and according to simulations (not shown) requires rather large pair-breaking scattering in order to partially fill the superconducting gap, we can not rule it out at the present stage.

\subsection{Results of band structure calculations}

\begin{figure*}[]
\includegraphics[width=\textwidth ]{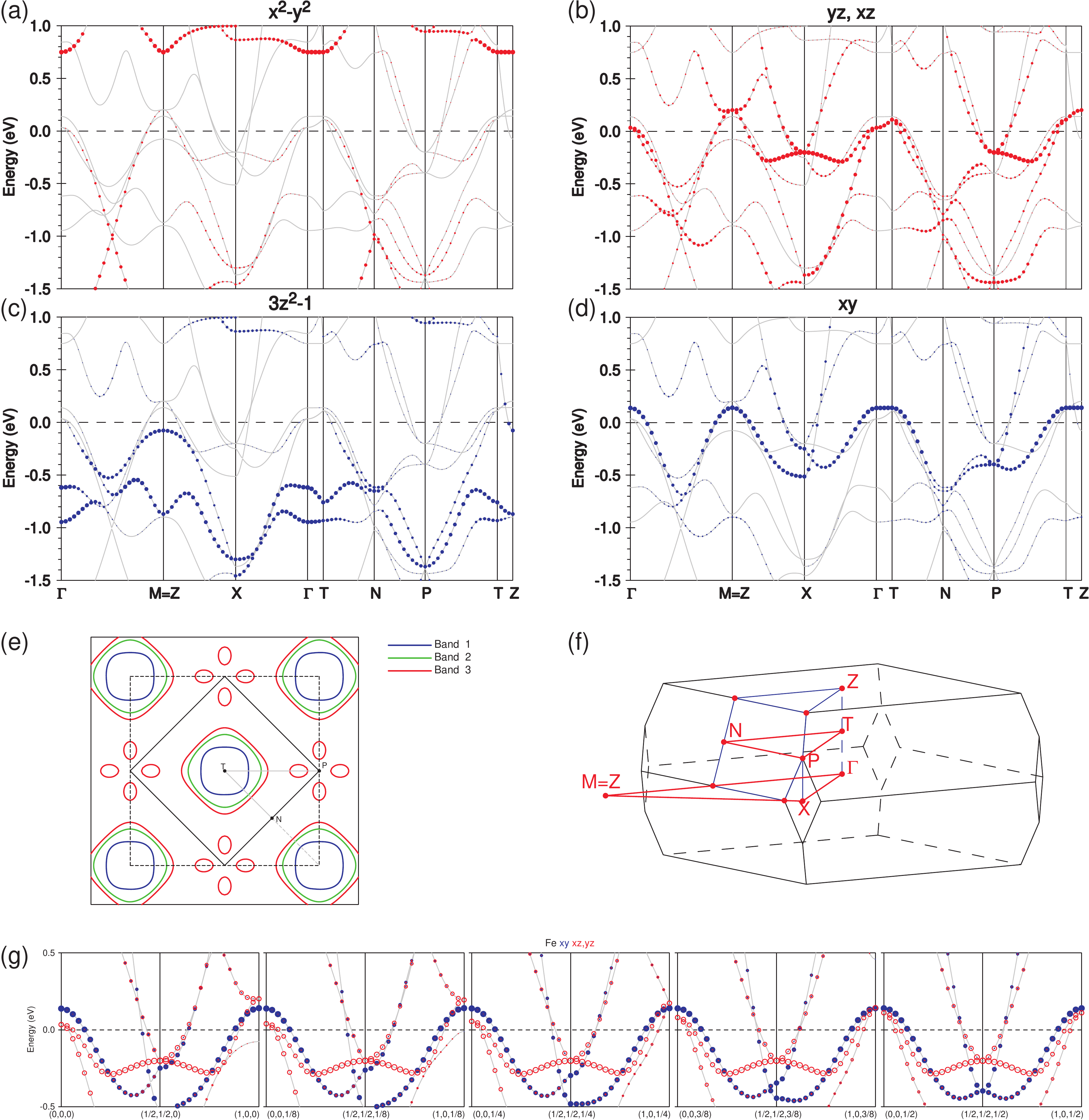}
\caption{Calculated band structure of BaFe$_2$As$_2$. Panels (a,b,c,d) show the band dispersion and contribution from different iron $3d$ orbitals. (e) The calculated band dispersion cut at 250\,meV below the Fermi level. (f) Three-dimensional Brillouin zone. (g) Band dispersion for different $k_z$ values.}
 \label{Yaresko}
\end{figure*}

The calculated band structure of undoped BaFe$_2$As$_2$ is presented in Fig.~S\ref{Yaresko}. The electronic states at the Fermi level are formed almost entirely of iron $3d$ orbitals. Panels (a)--(d) present band dispersion with superimposed circles, denoting the weight of the atomic-like states in the decomposition of the wave functions. Panel (e) presents the isoenergetic contours, obtained by cutting the calculated band dispersion at 250\,meV below the Fermi level. The similarity of the four hole-like elongated ellipses at the BZ corner to the propellers, observed in FS maps, measured by ARPES, hints that the calculated and measured band structures can be matched if one slightly shifts and bends the bands. An important remark that is to be done here is that for denoting the orbital composition of the bands we use the reference frame with $x$ and $y$ axes directed along the diagonals of a two-dimensional Fe$_2$As$_2$ unit cell, and $z$ directed perpendicular to the iron-arsenic layers (avoid confusion: momentum components $k_x$ and $k_y$ are directed parallel to the boundaries of Fe$_2$As$_2$ unit cell). The dispersion of the spectral weight near the BZ corner in the energy-momentum cut [see Fig.~\ref{ARPES_calc}~(c)] in turn reminds the dispersion of the $3d_{xz,yz}$ band, situated right below the bottom of the electron-like pockets in the calculation [see Fig.~S\ref{Yaresko}~(b)]. Another characteristic feature of the calculated band structure is a strongly three-dimensional band at $(k_x=0,k_y=0)$ coming from high above the Fermi level as a function of $k_z$ and starting to interact with two $3d_{xz,yz}$ bands at $k_z$ values of about $\pi\pm\pi/4$ [see Fig.~S\ref{Yaresko}~(a--d)]. As already mentioned above, such behavior is consistent with rather abrupt variation of many characteristics of the photoemission spectrum in the vicinity of some values of $h\nu$ and no considerable variation for large regions in between. An independent confirmation that peculiar $h\nu$ values of 33.5, 60 and 93\,eV correspond to the Z point comes from the fitting of the $h\nu$ dependence of the superconducting gap \cite{D_hv}. The detailed dependence of the electronic states of the calculated band structure on $k_z$ is shown in Fig~S\ref{Yaresko}~(g): at $(k_x=0,k_y=0)$ the interaction of the quasi two-dimensional bands, supporting $\Gamma$ FS sheets, with the mentioned band, coming from higher energies, results in the onset of the $k_z$ dispersion and variation of the orbital composition at $k_z=\pi$.

The matrix element of a transition from an initial state $u_i$ to a final state $u_f$ upon action of the electromagnetic field of the incoming light can be written as
\begin{equation}
\int u_f^*(\mathbf{r})\mathbf{A}\cdot\nabla u_i(\mathbf{r}) d^3\mathbf{r},
\label{int_fi}
\end{equation}
where integration is performed over the whole coordinate space. Zero matrix element would mean that the transition does not happen, i.e. some particular electronic states do not contribute to the observed photoemission signal. Symmetry considerations reveal many cases when the matrix element is zero: e.g., if there exists such a mirror plane that the integrand is odd with respect to the corresponding reflection, the whole integral is zero. It is particularly convenient to analyze parity with respect to the plane containing both the normal to the sample surface and the momentum of photoelectron flying to the analyzer. Such an analysis has been carried out for ARPES on iron
arsenides to identify the bands in the center of BZ, and in particular to resolve two closely located bands in the center of BZ \cite{FengPRL}. In Fig.~\ref{ARPES_calc}~(c,d) we propose a similar analysis for the energy-momentum cut passing through the BZ diagonal and cutting through all different FS sheets. For the case of the bands, formed by combination of $d_{xz,yz}$ orbitals, the orbitals lying in a plane are even with respect to reflection in it, while the ones standing ``perpendicular'' to the plane are odd. The most prominent difference upon switching polarization occurs for the band, composed of $d_{yz}$ orbitals in Fig.~\ref{ARPES_calc} (for the orbital character, obtained from the calculation, see Fig.~S\ref{Yaresko} and Fig.~S3~(e) in Ref.~\onlinecite{Feng_BFAP}).

It is still unclear, whether the entire propeller-shaped intensity, seen in ARPES (including central small electron-like pocket), can be satisfactory reproduced by cutting the calculated band structure at the suitable energy level and emulation of the self energy effects. Additionally, temperature dependence of ARPES signal from propellers has been observed \cite{VolodyaNature,Evtushinsky_JPSJ}, suggesting that appearance of propeller bands at the Fermi level has a connection to the fluctuations of magnetic order. Still, looking at the orbital composition of the calculated bands in this region [Fig.~S\ref{Yaresko}] and at the polarization dependence of ARPES spectra [Fig.~\ref{ARPES_calc}(c,d)], together with calculated and experimental dispersions it seems very reasonable to suggest that all propeller constituents, i.e. both shaft and blades, are ``assembled'' of iron $3d_{xz,yz}$ orbitals.

\begin{figure*}[]
\includegraphics[width=1\textwidth ]{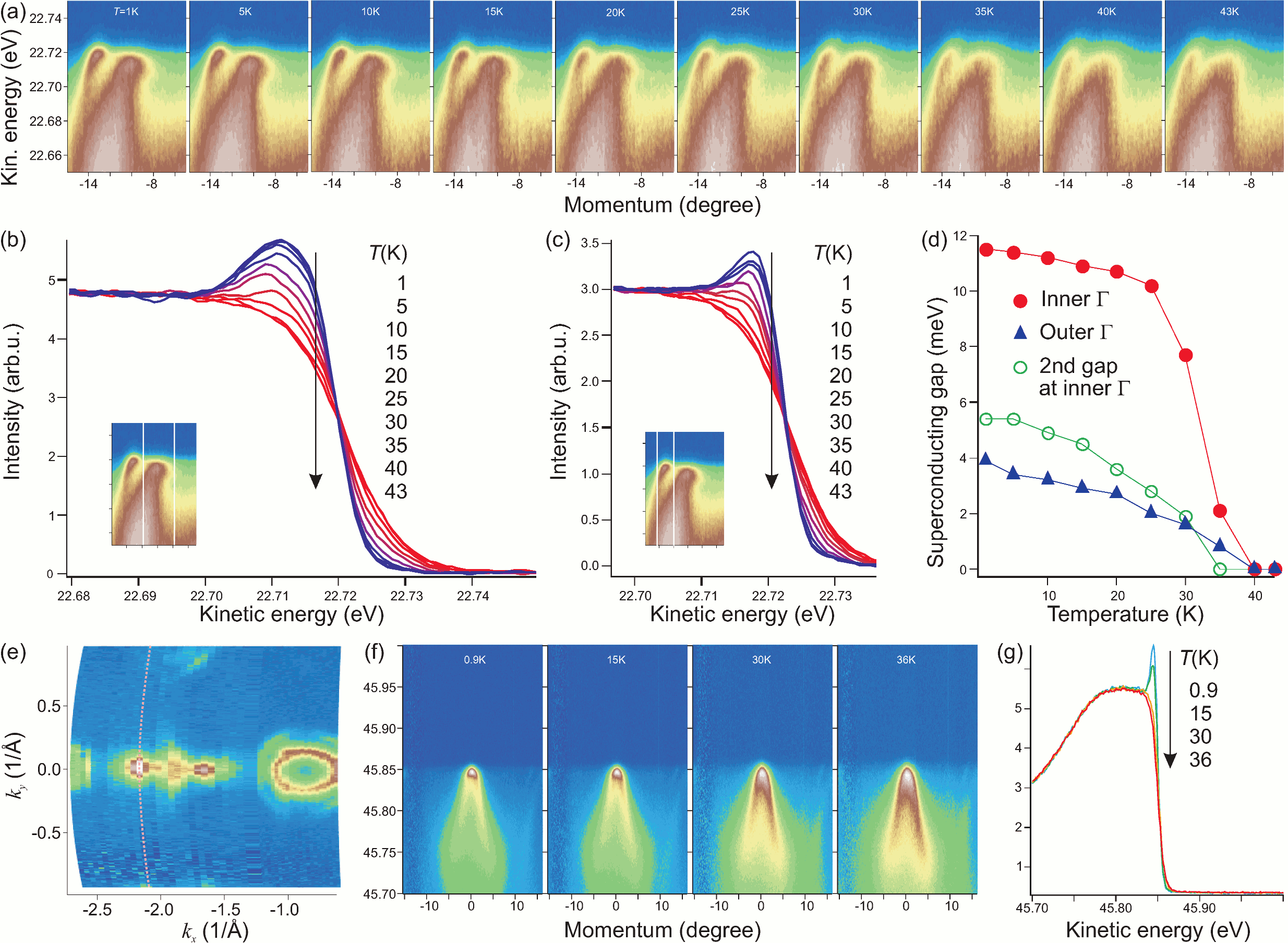}
\caption{(a) Sequence of temperature-dependent measurements, performed on Ba$_{1-x}$K$_{x}$Fe$_2$As$_2$ with $T_{\rm c}$ of 38\,K, for the energy-momentum cut, passing through the $\Gamma$-barrels in the range of 1--43\,K. The emergence of superconducting gap shows up as appearance of the ``beaks'' at Fermi crossings of both inner and outer $\Gamma$-barrels. (b) Temperature dependence of the EDC, referring to the inner $\Gamma$-barrel, integrated in the range shown in the inset. The spectrum above $T_{\rm c}$ is nothing but Fermi step, while the coherence peak starts to grow at superconductivity onset. (c) Same as (b), only for the outer $\Gamma$-barrel. (d) Temperature dependence of the superconducting gaps, derived from fit of integrated EDC to Dynes function. Panels (e,f,g) present data, taken from Ba$_{1-x}$Na$_{x}$Fe$_2$As$_2$ with $T_{\rm c}=34$\,K. (e) Fermi surface map. (f) Temperature dependence of the energy-momentum cut passing through the center of the propeller's blade, as shown by dashed line in panel (e). (g) Temperature dependence of the integrated EDC, revealing the appearance of prominent coherence peak below $T_{\rm c}$.}
 \label{gap_Tdep}
\end{figure*}

\begin{figure*}[]
\flushright
\begin{minipage}[l]{0.65\linewidth}
\includegraphics[width=\textwidth]{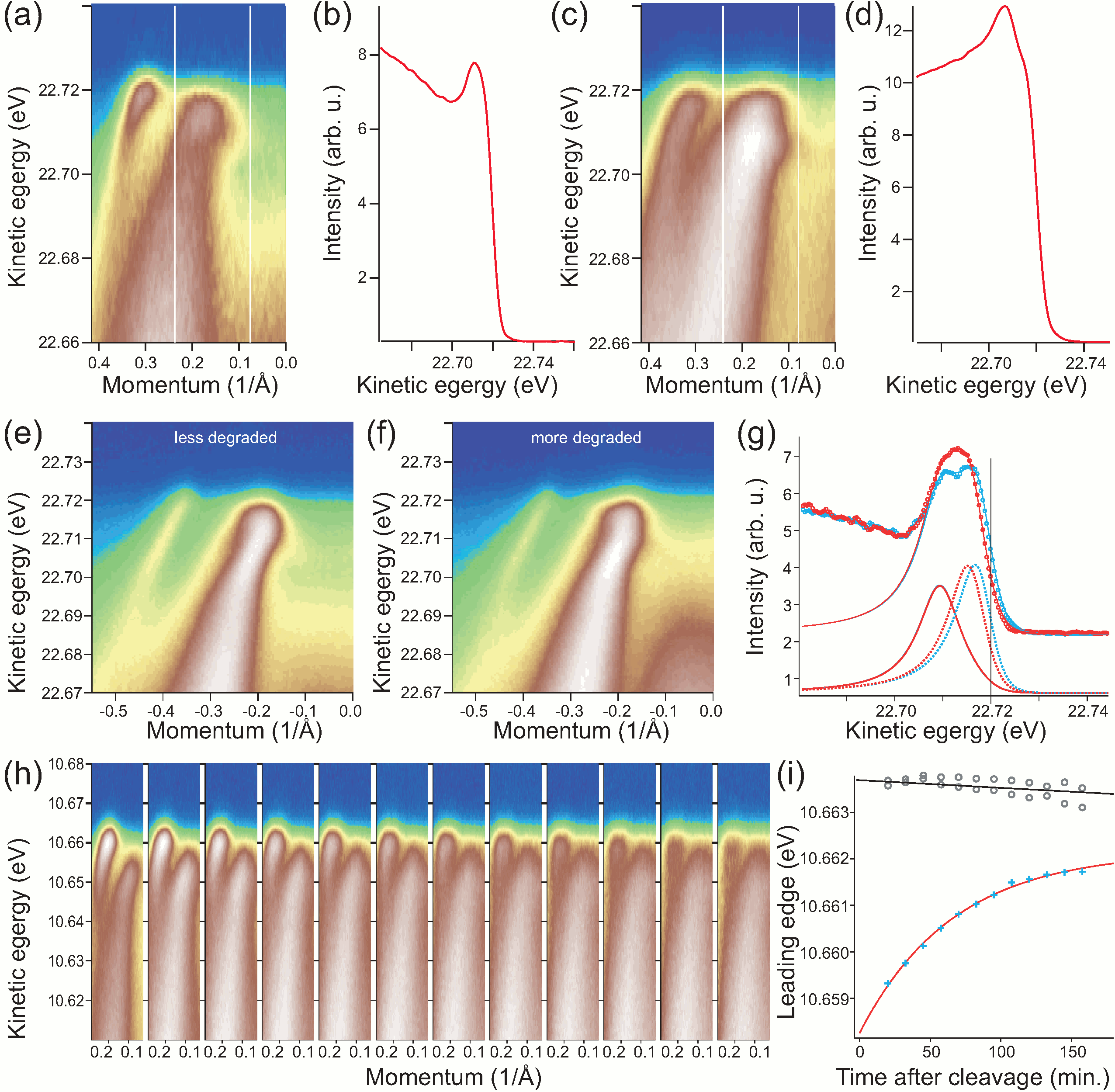}
%\caption{default}
%\label{fig:figure1}
\end{minipage}
\hspace{0.000\linewidth}
\begin{minipage}[r]{0.335\linewidth}
\caption{Deterioration of the sample surface with time after cleavage results in reduction of the gap in surface-related part of the photoemission spectrum. At the same time another, bulk, component of the signal exhibits a robust time-independent gap value. (a) Energy-momentum cut, passing through the $\Gamma$-point, recorded soon after cleavage. (b) EDC, integrated around the Fermi crossing of the inner $\Gamma$-barrel, as indicated in the panel (a). (c) The same cut, recorded from the same cleave after more than 25 hours. (d) Corresponding EDC. (d) Corresponding EDC. (e,f,g) is a different data set, where (e) and (f) panels represent signals from less and more deteriorated surface respectively. (g) Near-$k_{\rm F}$ EDCs after deconvolution together with fits to two lorentzians. The peak, corresponding to bulk signal remains unchanged, while the peak, corresponding to the surface signal with reduced gap shifts towards Fermi level. (h) Third data set: time evolution of the similar cut, passing through the $\Gamma$-barrels, recorded immediately after cleavage at residue pressure in the measuring chamber of $1.3\cdot10^{-10}$\,mbar and using Janis ST400 cryomanipulator. (i) Time dependence of the leading edge position (LEG) for the outer (grey circles) and inner (blue crosses) $\Gamma$-barrels.}
\label{nonSC}
\end{minipage}

\end{figure*}

\subsection{Temperature dependence of the superconducting gap}

In Fig.~S\ref{gap_Tdep} we show the temperature dependence of the ARPES spectra for temperature range from above $T_{\rm c}$ down to 1\,K. As one can see from Fig.~S\ref{gap_Tdep}~(a,b,c), spectral signatures of the superconducting gap\,---\,``beaks'' at Fermi crossings in the energy-momentum cuts and coherence peak in the integrated spectrum\,---\,appear when cooling through $T_{\rm c}$. These signatures are present for both outer and inner $\Gamma$-barrels, implying that both large gap on the inner $\Gamma$-barrel and the small gap on the outer $\Gamma$-barrel, open at $T_{\rm c}$. As discussed in the next section of the Supplemetary Materials, ARPES spectra of BKFA contain large contribution of a component with reduced superconducting gap, stemming from the surface layer. This component of the photoemission data is taken into account upon fitting the data \cite{non_BCS_spectr}, and the magnitude of the reduced gap can be extracted separately. In Fig.~S\ref{gap_Tdep}~(c) are shown temperature dependencies for the gap magnitude on the outer $\Gamma$-barrel, two unresolved inner $\Gamma$-barrels and surface component of the inner $\Gamma$-barrels. Thus, both small gap on the outer $\Gamma$-barrel and large gap on the inner $\Gamma$-barrels emerge upon entering the superconducting state, and consequently are unambiguously related to the superconductivity.

Fig.~S\ref{gap_Tdep}~(e) shows the Fermi surface map of optimally doped Ba$_{1-x}$Na$_{x}$Fe$_2$As$_2$ with $T_{\rm c}$ of 34\,K. As one can see, the Fermi surface of Ba$_{1-x}$Na$_{x}$Fe$_2$As$_2$ consists of hole-like $\Gamma$-barrels in the BZ center and propeller-like structure at the BZ corner, illustrating the already mentioned similarity of the electronic structures of sodium- and potassium-doped BaFe$_2$As$_2$ (see also Ref.\,\onlinecite{BNFA2}). The temperature dependence of the cut, passing through the propeller's blade [pink dashed line in panel (e)] is
presented in Fig.~S\ref{gap_Tdep}~(f). The hole-like dispersion of the band, supporting propeller's blades, is clearly visible. The development of the superconducting gap can be recognized as modification of the spectral function in the vicinity of the Fermi level in panel (f) and by growth of the prominent coherence peak in the integrated spectrum in panel (g). Clear observation of the large superconducting gap implies that the propeller-like Fermi surface takes active part in the superconductivity.

\subsection{Additional component of the photoemission signal from the inner $\Gamma$-barrel}

Already in first studies of the superconducting gap in BKFA an additional component of the inner $\Gamma$-barrel was noticed and attributed to the non-superconducting part of the photoemission spectrum \cite{DingEPL, EvtushinskyPRB}. More recent experiments have shown that this feature is not entirely non-superconducting, but bears a reduced superconducting gap of 3--7\,meV \cite{ShinScience, FengPRL}. To investigate this issue more thoroughly we have performed time-dependent measurements of the intensity distribution in the vicinity of the Fermi level in the superconducting state. Fig.~S\ref{nonSC}~(a--d) shows two equivalent spectra taken from the same cleaved sample surface soon after cleavage and on the next day. Aged sample surface obviously exhibits a prominent non-superconducting (or bearing very small gap) component, while for the fresh cleavage the secondary component is gapped with $\sim6$\,meV gap, and therefore, is not that prominent, merging visually with the peak from the $\sim11$\,meV gap. Interestingly, both in case of aged and fresh surfaces, the large gap remains the same, $\sim11$\,meV, i.e. is robust with respect to surface degradation. The inner $\Gamma$-barrel is double-walled, so generally one might expect an additional feature at this location in the Brillouin zone: each of these two nearly degenerate bands might bear different superconducting gaps. However, such explanation for two distinct gaps in this momentum region faces the following difficulties: (i) the two inner $\Gamma$-barrels are highly degenerate and have virtually the same orbital composition of the wave functions, therefore it seems unlikely that they bear such different gap magnitudes (6 vs. 10.5\,meV), and (ii) one of the gaps almost completely vanishes with time, while the other remains unchanged, also implying different origin of the two neighboring features, while the impact of surface degradation on the two components of the inner $\Gamma$-barrel is expected to be very similar as they both are composed of the same iron $d_{xz,yz}$-orbitals. A more plausible explanation is that the spectral feature with reduced gap is connected to the photoemission signal from the topmost surface layer, and consequently, should be filtered out when speaking of the bulk gap structure.

Fig.~S\ref{nonSC}~(e,f) show two spectra, taken in the very same conditions from the same cleave even with no adjustment of sample position. The difference between these two spectra is due to surface degradation, related to the residual pressure in the measuring chamber. Visually presence of a component with reduced gap is seen as a ``hat'' on top of the feature bearing large gap [see Fig.~S\ref{nonSC}~(c,f)]. The difference between spectra in panels (e) and (f) is that this hat is shifted more towards the Fermi level, i.e. further from the feature with the robust large gap. Panel (g) shows a detailed analysis of the near-$k_{\rm F}$ EDC from panels (e) and (f). The peak in the EDC is fitted to two Lorentzians, also shown in panel~(g) below the data. Remarkably, the difference in the fits for more and less degraded spectra is only in the shift of the Lorentzian, located at the lower binding energy to even lower; at the same time the position of the Lorentzian at higher binding energy and relative weights of Lorentzians remain unchanged. This allows us to state that the main effect of surface degradation is in further closing of the reduced gap.

To get some quantitative estimate of the residual pressure effect on the surface degradation, we have performed a time-dependent measurements of the same spectrum capturing $\Gamma$-barrels [Fig.~S\ref{nonSC}~(h)]. The residual pressure was $1.3\cdot10^{-10}$\,mbar, and time interval between consequent measurement, shown in panel (h), was 12.5\,minutes. In the Fig.~S\ref{nonSC}~(i) we plot the temperature dependence for positions of the leading edge midpoints of the inner and outer $\Gamma$-barrels. The time evolution for the inner $\Gamma$-barrel shows exponential behavior with time constant of approximately one hour. The same temperature dependence for the outer $\Gamma$-barrel shows small decrease instead of expected small increase due to onset of different effects (variation of intensity distribution, broadening of spectral features etc.) influencing the leading edge midpoint position.

%Although the two components of the inner $\Gamma$-barrel are not resolved in ARPES data in the vast majority of cases, revealing itself rather as a broadening of the spectra (especially noticeable once one compares MDC widths for inner and outer $\Gamma$-barrels), at some experimental conditions, using suitable excitation energy, one can clearly distinguish all three FS sheets around $\Gamma$-point. Particularly it is possible at $h\nu$=33\,eV, where at the same time the $h\nu$-dependence of the large gap has a cusp. At this $h\nu$ the splitting of two inner $\Gamma$-barrels is the largest, while the gaps still appear to be quite close, 5.5 and 4.5\,meV.

We emphasize once more that the large superconducting gap on the inner $\Gamma$-barrel, is a robust property of the ARPES spectra of optimally doped BKFA, while the reduced gap is strongly dependent on the experimental conditions, and most likely is related to the surface. It is worthwhile noting that in the case of ARPES experiments on LiFeAs no additional components of the signal with reduced gap has been notices and the spectra generally deteriorate less with time passed after cleavage \cite{BorisenkoLiFeAs, DimaLiFeAs}. The reason is that, in contrast to 122 iron arsenides, LiFeAs has a natural cleavage plain and topmost superconducting Fe-As layer appears to be covered with a whole Li layer.

Presence of two components in the spectra\,---\,the one with a reduced gap value, dependent on the surface quality, and the other with large robust gap\,---\,imply that the observed spectra represent a sum of a signal coming from the topmost surface layer and a signal from the deeper subsurface layers. The fact that the difference between the surface and subsurface signal manly consists in the different gap values, while no difference in the band dispersion was detected, suggests that surface effects are not crucial for determination of the band dispersion from ARPES on Ba$_{1-x}$K$_{x}$Fe$_2$As$_2$. Together with matching Hall measurements \cite{Evtushinsky_JPSJ} and with opening of the large superconducting gap at propellers [Fig.~S\ref{gap_Tdep}~(f)], it gives good substantiation for the bulk origin of the propeller-like Fermi surface. Large superconducting gap opening at propeller implies active role in superconductivity.


\begin{thebibliography}{00}
\bibitem{GBarrel} Though generally we consider essentially three-dimensional electronic structure of BKFA in this Manuscript, for the hole-like sheets in the BZ center we will use the notation of ``$\Gamma$-barrels'', stemming from two-dimensional terminology. We will also sometimes refer to the ``inner'' and ``outer'' $\Gamma$-barrels, implying that splitting of the two inner FS sheets is not principal for particular considerations.
\bibitem{BFAP_node} The behavior of the superconducting gap, suggesting presence of the gap nodes along $k_z$ on the $d_{xz, yz}$ $\Gamma$-barrel, was found for the case of  BaFe$_2$As$_{2-2x}$P$_{2x}$ \cite{Feng_BFAP_node}. The $k_z$-dependence of the gap is very reminiscent of the one, observed for Ba$_{1-x}$K$_{x}$Fe$_2$As$_2$, only with a deeper minimum at Z point. This is perfectly consistent with proposed here orbital dependence of the gap: in BaFe$_2$As$_{2-2x}$P$_{2x}$ the electronic states at $\Gamma$ point exhibit stronger dependence on $k_z$.
\bibitem{122KNa} The electronic structures of K- and Na-doped BaFe$_2$As$_2$ are very similar. ARPES spectra of both compounds are characterized by propeller-like FS, similar magnitude and momentum distribution of the superconducting gap including very similar $k_z$-dependnce. Possible differences between Ba$_{1-x}$K$_x$Fe$_2$As$_2$ and Ba$_{1-x}$Na$_x$Fe$_2$As$_2$ are insignificant for the presented here analysis.
\end{thebibliography}
\begin{thebibliography}{00}
\bibitem[S5]{D_hv} The fitting curves, used for approximation of the excitation energy dependence of the superconducting gap magnitudes on the $\Gamma$-barrels in Fig.~\ref{gap_kz}~(d), are $\Delta_{\rm large}(h\nu) = 10.5-5\cdot\sin^8(0.5\cdot6.7 \cdot \sqrt{0.262\cdot(h\nu+7.6)})$, $\Delta^{\prime}_{\rm large}(h\nu) = 10.5-6\cdot\sin^8(0.5\cdot6.7 \cdot \sqrt(0.262\cdot(h\nu+7.6)))$, $\Delta_{\rm small}(h\nu)=3.5$. Apart from the value of 7.6\,eV for the inner potential, from fitting of the data we, can infer that the peculiar points in the $h\nu$ dependence correspond to the $k_z=\pi$. i.e Z point in the threedimensional BZ. The deviation of the model from experimental data at photon energies below 20\,eV is expected from break down of the free electron approximation for the final state.
\end{thebibliography}
\end{document}